\documentclass[msmath,amssymb,aps,prl,twocolumn,epsfig,showpacs,bibliography,
lengthcheck,superscriptaddress]{revtex4-1}

\usepackage{graphicx}
\usepackage{dcolumn}
\usepackage{bm}
\usepackage{hyperref}
\usepackage{epstopdf}
\usepackage{subcaption}
\usepackage{caption}
\usepackage{footmisc}
\usepackage{amsmath}
\usepackage{color}
\begin{document}
\title{Search for monopole-dipole interactions at the sub-millimeter range with a $^{129}$Xe-$^{131}$Xe-Rb comagnetometer}
\author{Y.-K. Feng}
\author{D.-H. Ning}
\author{S.-B. Zhang}
\author{Z.-T. Lu}
\email{ztlu@ustc.edu.cn}
\affiliation{School of Physical Sciences, University of Science and Technology of China, Hefei 230026, China}
\author{D. Sheng}
\email{dsheng@ustc.edu.cn}
\affiliation{Department of Precision Machinery and Precision Instrumentation, Key Laboratory of Precision Scientific Instrumentation of Anhui Higher Education Institutes, University of Science and Technology of China, Hefei 230027, China}

\begin{abstract}
Monopole-dipole interactions involving scalar couplings between a spin and a massive particle violate both P- and T-symmetry, and can be mediated by axions. We use a $^{129}$Xe-$^{131}$Xe-Rb atomic cell comagnetometer to measure the ratio of precession frequencies between the two xenon isotopes, and search for changes of the ratio correlated with the distance between the atomic cell and a non-magnetic  bismuth germanate (BGO) crystal. A modulated Rb polarization scheme is used to suppress systematic effects by two orders of magnitude. The null results of this search improve the upper limit on the coupling strength $g_{s}^Ng_{p}^{n}$  over the interaction range 0.11 - 0.55~mm, and by a maximum improvement factor of 30 at 0.24~mm. The corresponding propagator mass range of this new excluded region covers 0.36 - 1.80 meV.
\end{abstract}

\maketitle

Axion, a hypothetical particle to solve the strong CP problem in QCD~\cite{peccei1977}, also provides a possible source for the cold dark matter in the Universe~\cite{ipser1983}. Numerous methods have been employed to search for axions~\cite{sikivie2021,rpp2020}, and most laboratory efforts are based on axion-photon conversions in the presence of a magnetic field~\cite{ADMX2010,ADMX2018,ADMX2020,cast2005,cast2017,cameron1993,PVLAS2006,OSQAR2015}.
In the past decade, table-top experiments searching for anomalous forces between atoms and macroscopic objects have emerged to provide complementary methods to search for new physics beyond the Standard Model~\cite{safronova2018,terrano2021}.

One type of the anomalous forces, the monopole-dipole interaction between particles $a$ and $b$, can be expressed as~\cite{moody1984,dobrescu2006,fadeev2019}:
\begin{equation}
\label{eq:Vr}
V(\boldsymbol{r})=\hbar^{2} g_{s}^{a} g_{p}^{b} \frac{\boldsymbol{\sigma}_{b} \cdot \hat{\boldsymbol{r}}}{8 \pi m_{b}}\left(\frac{1}{r \lambda}+\frac{1}{r^{2}}\right) e^{-\frac{r}{\lambda}}=v(r)\hat{\boldsymbol{r}}\cdot\boldsymbol{\sigma}_{b},
\end{equation}
where ${g_s}$ and ${g_p}$ are the scalar and pseudoscalar coupling strength, respectively. $m$ is the particle mass, $\hat{\boldsymbol{r}}$ is the unit vector connecting the two particles, $\hbar\bm{\sigma}/2$ is the particle spin, and $\lambda$ is the interaction length, which is also the reduced Compton wavelength of the interaction propagator. Due to the pseudoscalar coupling term $\boldsymbol{\sigma}_{b} \cdot \hat{\boldsymbol{r}}$, monopole-dipole interactions violate both P- and T-symmetry and can be mediated by axions. This interaction can also be viewed as the coupling between the spin of particle $b$ and an effective magnetic field generated by particle $a$, with the field amplitude proportional to $v(r)$ in Eq.~\ref{eq:Vr} and the field direction along $\hat{\boldsymbol{r}}$. In this way, atomic magnetometers and comagnetometers are suitable tools for the measurement of this exotic coupling field.

In this paper, we focus on the couplings between nucleons and neutron spins. For the classical axion window (1 $\mu$eV $< m_a <$1 meV, 0.2 mm $< \lambda <$ 0.2 m)~\cite{bradley2003}, the laboratory upper limits on the coupling strength $g_s^Ng_p^n$ are set by works using nuclear spin comagnetometers~\cite{bulatowicz2013,tullney2013} and self-compensating comagnetometers~\cite{lee2018}, and works studying the $^3$He depolarization time~\cite{guigue2015}. Moreover, efforts exploring resonant excitations in atomic systems~\cite{Arvanitaki2014} are under development to bridge the gap between constraints set by laboratory works and these set by analyses of astronomical events~\cite{raffelt2012}.

In this work, we search for the monopole-dipole interactions between neutron spin from a $^{129}$Xe-$^{131}$Xe-Rb comagnetometer and nuclei of a non-magnetic BGO crystal at the interaction length 0.1 - 0.6~mm. This comagnetometer system has been used in Ref.~\cite{bulatowicz2013} to search for the monopole-dipole interactions. In that experiment, the magnetic field generated by the polarized Rb atoms was identified as the dominant systematic effect. Here, we employ a modulated Rb polarization scheme~\cite{limes18}, and suppress the systematic effect from polarized Rb atoms by two orders of magnitude. Moreover, the measurement sequence of this work is designed to further suppress residual effects from imperfect controls of experiment parameters, by periodically switching three experiment conditions: the bias field direction, the pumping beam polarization, and the position of the BGO mass. The results from this work improve the upper limits of the coupling strength $g_{s}^Ng_{p}^{n}$  in the interaction range 0.11 - 0.55 mm.

%

\begin{figure}[hbt]
\includegraphics[width=3in]{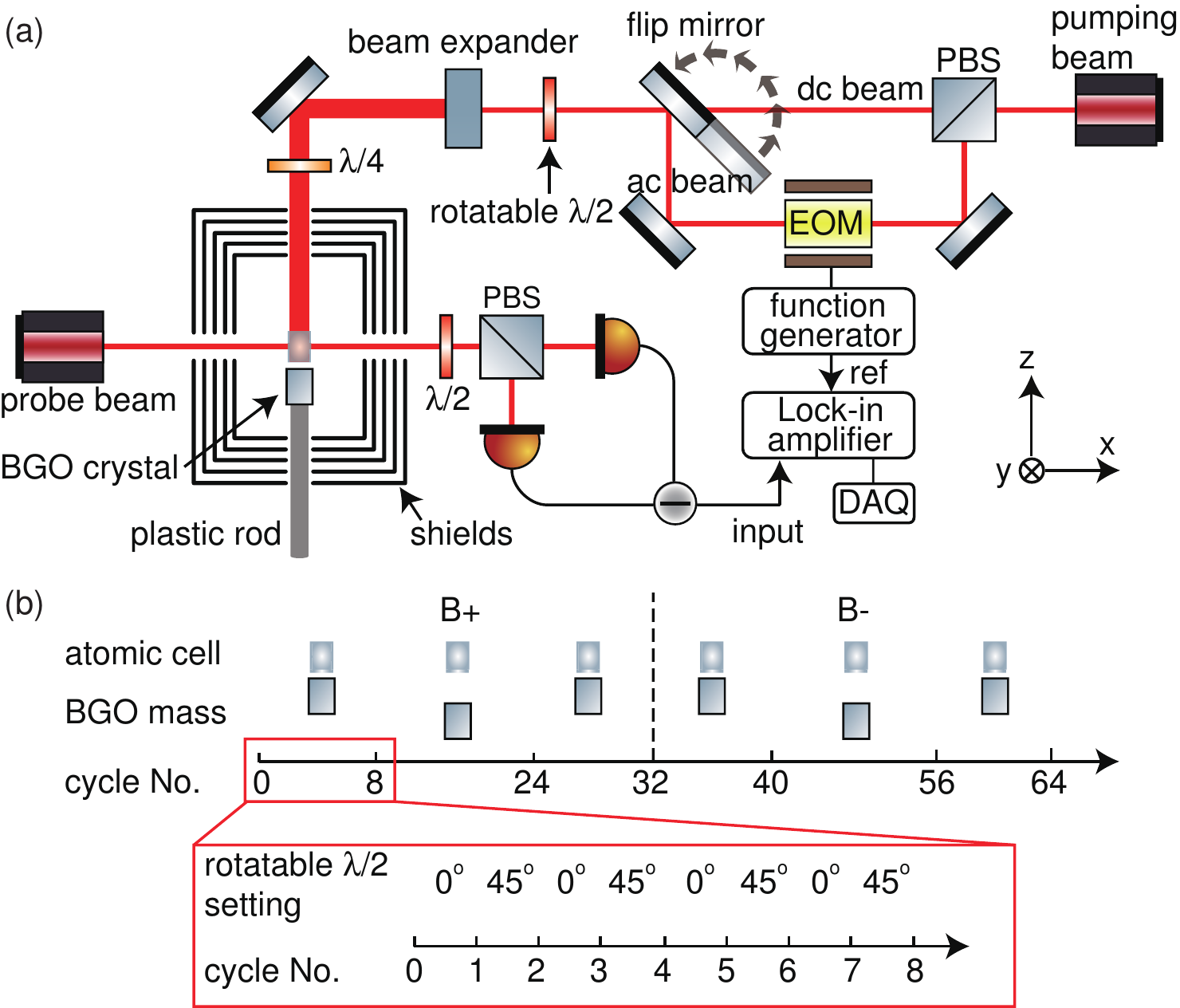}
\caption{\label{fig:apparatus}(Color online) (a) Experiment setup. (b) Three types of switches are made in a single measurement round: the bias field direction (B$_+$ and B$_-$), the pumping beam polarization (corresponding to the 0$^\circ$ and 45$^\circ$ angle setting of the rotatable half-wave plate), and the position of the BGO mass (near and far).}
\end{figure}

A $^{129}$Xe-$^{131}$Xe-Rb comagnetometer is used to take advantage of large spin-exchange collision rates~\cite{walker1997} and similar collision properties~\cite{bulatowicz2013,feng2020} between Xe isotopes and Rb.  The comagnetometer cell is a rectangular Pyrex glass cell with an inner dimension of 4$\times$4$\times$9 mm$^3$, and its end wall perpendicular to the long axis has a thickness of 0.5 mm. This asymmetric cell shape is chosen to amplify the quadrupole splitting so that the $^{131}$Xe spectral lines are resolved~\cite{wu1987,feng2020}. The cell contains Rb in natural abundances, 4 Torr of $^{129}$Xe, 15 Torr of $^{131}$Xe, 8 Torr of H$_2$~\cite{kwon81,wu1990} and 400 Torr of N$_2$. Figure~\ref{fig:apparatus}~(a) shows the main experiment setup. The cell is mounted inside an oven with the temperature controlled at around 105 $^{\circ}$C, and the oven is placed in the center of a magnetic shield structure with five layers of mu-metal shields.  A  bias field points along the $z$ axis inside the shields, which also coincides with the east-west direction, so that the comagnetometer is insensitive to the earth rotation~\cite{bfmisc}.

As shown in Fig.~\ref{fig:apparatus}~(a), the optical pumping beam is divided into two beams by a polarization beam splitter (PBS), the dc and ac beam, each with the same power of 20 mW and the same diameter of 1 cm. Its frequency is tuned to the resonance of the Rb D1 line. The polarization of the dc beam is constant over time, while the polarization of the ac beam is modulated by an electric-optical modulator (EOM). A flip mirror is used to send either the dc or ac beam to the cell. The polarizations of both dc and ac beams are also controlled by a rotatable half-wave plate. The probe beam has a diameter of 1~mm, a power of 3.5~mW, and a blue detuning of 20 GHz from the Rb D1 line. It passes through the cell along the $x$ direction to probe the Rb magnetometer signal. The probe beam is initially linearly polarized and, after transmitting through the cell, the direction of its polarization is analyzed by a polarimeter.

BGO is chosen as the monopole material due to its  non-magnetic property, high density (7.12~g/cm$^3$), and low thermal conductivity. The cylindrical BGO mass has a diameter of 1~cm, a length of 2~cm, and a nucleon number density of 4.5$\times$10$^{24}$/cm$^3$. It is held by a plastic rod, and placed coaxially with the atomic cell. A translation stage is used to move the BGO, so that the minimum distance between the front surface of the BGO and the inner surface of the cell is 0.74$\pm$0.02 mm, and their maximum distance is 3.25~mm. A camera is used to monitor and measure the movement of the BGO mass. In the experiment, we focus on the ratio $R$ between precession frequencies of the two Xe isotopes correlated with the BGO mass positions. $R$ can be expressed as
\begin{equation}
R=\frac{\omega_{129}}{\omega_{131}}=\frac{\gamma_{129}B+X_{129}}{\gamma_{131}B+X_{131}},
\end{equation}
where $\omega_i$ is the magnetic dipole precession frequency of the isotope $^i$Xe, $\gamma_i$ is the corresponding nuclear gyromagnetic ratio, and $X_i$ is the frequency shift of $^i$Xe due to exotic interactions.

A measurement round consists of 64 cycles (Fig.~\ref{fig:apparatus}~(b)), with each cycle being 160 s long.  For the first 60 s, the dc pumping beam with a $\sigma+$ polarization enters the cell, and the Xe atoms are hyper-polarized via spin-exchange collisions with the polarized Rb atoms~\cite{walker1997}. As the dc-pump period concludes, a short $\pi/2$ pulse ($\sim$0.1 s long) for both Xe isotopes is applied in the $y$ direction to tilt the Xe polarizations into the $xy$ plane. Meanwhile, the flip mirror is switched to reflect the ac pumping beam into the cell. The polarization of the ac pumping beam is modulated between $\sigma+$ and $\sigma-$ at the frequency of 198 Hz, which is much higher than the Larmor frequencies of Xe isotopes, yet still lower than the optical pumping rate of  Rb atoms. In this way, the time-averaged effective magnetic field B$_{Rb}$ generated by the polarized Rb atoms is reduced by two orders of magnitude, resulting in significant suppression of systematic effects. Also reduced is the field gradient of B$_{Rb}$, resulting in longer depolarization times of Xe~\footnote{See Sec.~I of the supplementary materials}. For the next 90 s, while the ac pumping beam is on, the Xe precession signals are continuously recorded by the Rb magnetometer. Once the precession data recording concludes, we spend 10 s to rotate the half-wave plate by 45 $^\circ$, in order to switch the circular polarizations of both the dc and ac pumping beams. The measurement cycle is then repeated with identical parameters, but with the circular polarizations of the pumping beams reversed. The results of the two-cycle pair, the $\sigma+$ cycle and $\sigma-$ cycle, are averaged to generate a value of the frequency ratio $R$. This average helps to reduce the residual effects from the imperfect $\pi/2$ pulse and asymmetrical modulations in the ac pumping beams.

During an entire measurement round that consists of 64 cycles, the BGO mass is moved to search for any correlation effects, and the bias B field direction is switched to further reduce systematic effects. In each round, the bias field is pointed along the $+z$ direction during the first 32 cycles, and to the $-z$ direction in the next 32 cycles. During this time, the BGO mass is adjusted between the “near” and “far” positions as shown in Fig.~\ref{fig:apparatus}~(b). The frequency ratio difference due to the BGO mass movement obtained in the first 32 cycles can be expressed as
\begin{equation}
\delta R_+=\bar{R}_{B+,near}-\bar{R}_{B+,far}.
\end{equation}
The corresponding difference $\delta{R}_-$ can be extracted for the next 32 cycles. Finally, the deviation extracted from one complete measurement round is
\begin{equation}~\label{eq:dx}
\Delta X=\Delta X_{129}-R\Delta X_{131}=\frac{\delta R_+-\delta R_-}{2}\omega_{131},
\end{equation}
where $\Delta X_i$=$X_{i,near}-X_{i,far}$.

\begin{figure}[hbt]
	\includegraphics[width=3in]{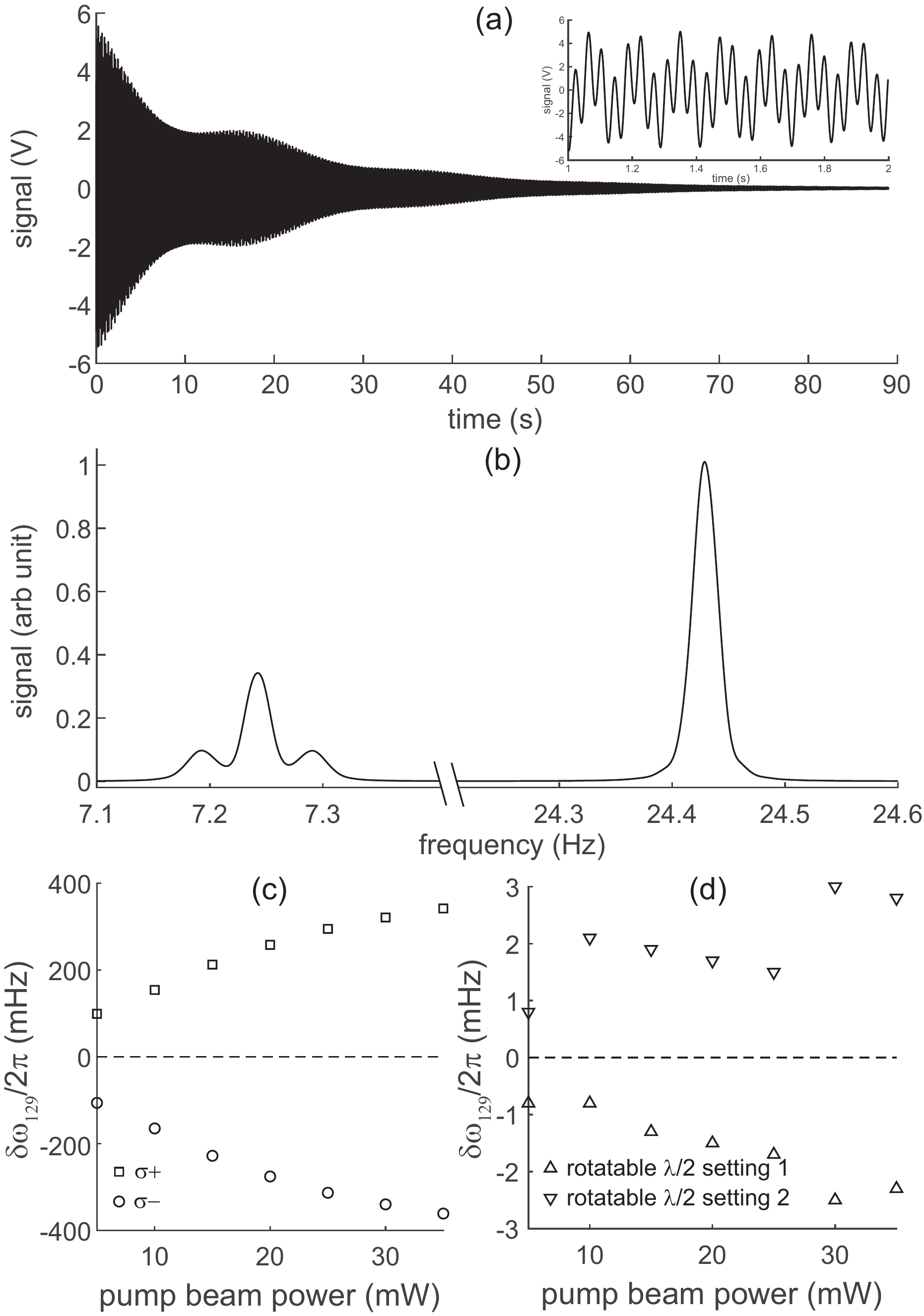}
	\caption{\label{fig:data} (a) A typical precession signal of Xe atoms at a bias field of 20.6 mG, with an amplified view of the data in the inset. (b) The amplitude spectrum of the signal in plot~(a). (c) $\delta\omega_{129}$, deviation of $\omega_{129}$ as a function of the pumping beam power, when probed by Rb magnetometers using dc pumping beams with different pump polarizations (square and circle points). (d)  $\delta\omega_{129}$ as a function of the pumping beam power, when probed by modulated Rb magnetometers with different rotatable half-wave plate settings (triangle points) as illustrated in Fig.~\ref{fig:apparatus}~(b).  The horizontal dashed lines in plot~(c) and (d) correspond to $\delta\omega_{129}/2\pi=0$~Hz.}
\end{figure}

A typical set of precession data with a bias field of 20.6 mG is shown in Fig.~\ref{fig:data}~(a).  Its amplitude spectrum displays four peaks (Fig.~\ref{fig:data}~(b)). The peak around 24.4~Hz, with its frequency equal to $\omega_{129}/2\pi$, is due to the magnetic dipole precession frequency of $^{129}$Xe; the other three peaks near 7.2 Hz form the quadrupole-split spectrum of $^{131}$Xe, with the frequency of each peak corresponding to $\omega_{131,(-,0,+)}/2\pi$ in the ascending order. We use multiple exponential-decay-oscillation functions to fit the data:
\begin{equation}
\label{eq:Fit}
y=\sum_n A_{n}\sin\left[\omega_{n} (t-t_0)+\phi_{n}\right] e^{-(t-t_0)/T_{n}}+b,
\end{equation}
where the sum is over the four precession frequencies mentioned above, and the typical fitting errors of these frequencies are at the $\mu$Hz level. As discussed in Ref.~\cite{feng2020}, we use $\omega_{131}=\omega_{131,0}$ to extract the magnetic dipole precession frequency of $^{131}$Xe, because it brings in the largest signal-to-noise ratio in the data analysis. We compare the results with those choosing  $\omega_{131}=(2\omega_{131,0}+\omega_{131,+}+\omega_{131,-})/4$, and confirm that both results are consistent with each other.

The systematic effects due to B$_{Rb}$ are related with the small difference in the collision properties between Rb atoms and the two Xe isotopes~\cite{bulatowicz2013}. Such effects are studied by examining the influence of B$_{Rb}$ on the precession frequency $\omega$ of each Xe isotope. Since the Rb polarization depends on the pumping beam power $P$, we study the effect of $P$ on the deviation of $\omega_{129}$. The deviation can be as large as $2\pi\times0.4$~Hz when the Rb polarization is not modulated, and this deviation changes sign when the pumping beam polarization is reversed (Fig.~\ref{fig:data}~(c)). On the other hand, when the modulated Rb polarization scheme is applied, $\delta{\omega_{129}}$ is suppressed by two orders of magnitude (Fig.~\ref{fig:data}~(d)). In conclusion, this modulation method significantly suppresses related systematic effects in the comagnetometer.

\begin{figure}[hbt]
	\includegraphics[width=3in]{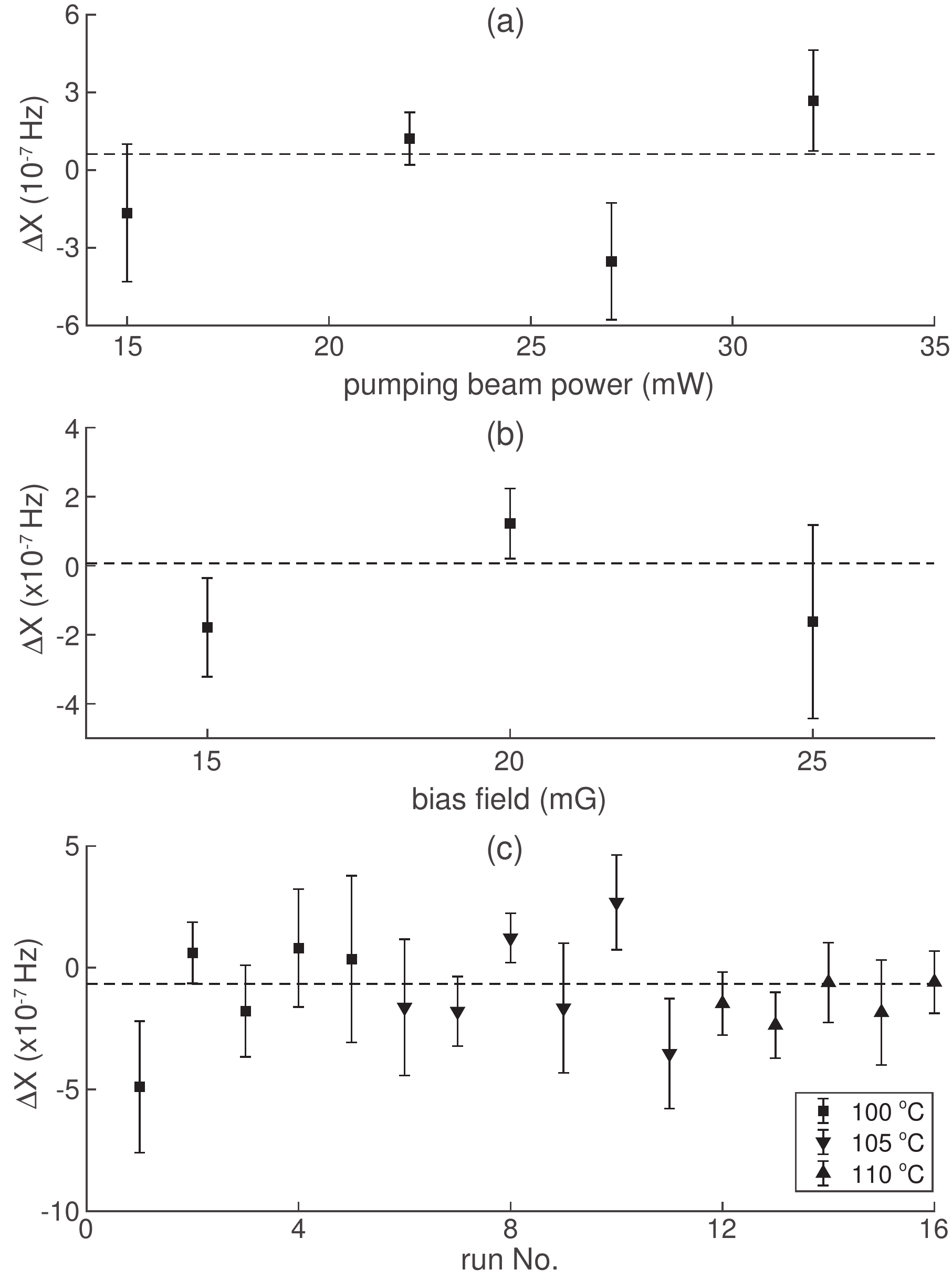}
	\caption{\label{fig:x} (a) $\Delta X$ is found to be independent of the pumping beam power; (b) $\Delta X$ is found to be independent of the bias field strength. Each data point corresponds to the average result of about 30 measurement rounds, and all the data are taken at the cell temperature of 105$^\circ$C. (c) Results collected over the course of two months. The reduced $\chi^2$ for this set of data is 1.03. The dashed line in each plot corresponds to the weighted average result of the demonstrated data.}
\end{figure}

Checks on systematic effects are performed by examining the dependence of $\Delta{X}$ on the pumping beam power (Fig.~\ref{fig:x}~(a)) and the bias field strength (Fig.~\ref{fig:x}~(b)). In addition, data are also collected while the cell temperature is varied by $\pm$5 $^\circ$C. Within the statistical errors, $\Delta{X}$ is found to be independent on these parameters. Figure~\ref{fig:x}(c) shows all the data acquired with different experiment conditions over a time span of two months. The weighted average result is $\Delta X/2\pi=-66\pm42$~nHz, leading to an upper limit of $|\Delta X/2\pi|<135$~nHz at the 95\% C.L..

\begin{figure}[hbt]
	\includegraphics[width=3in]{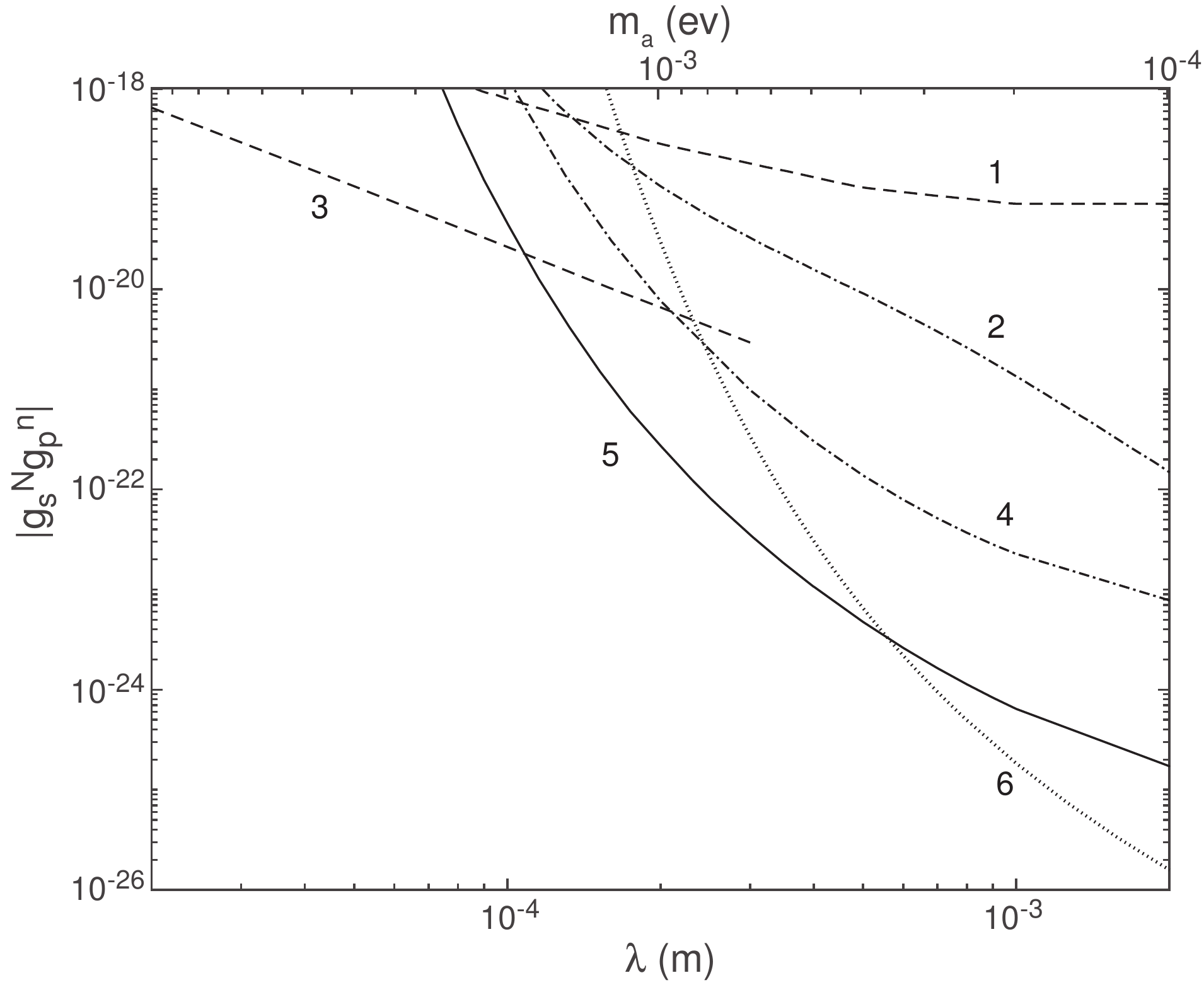}
	\caption{\label{fig:gsgn} Plot of constraints (95\% C.L.) on $|g_s^Ng_p^n|$ from different experiment efforts. Line 1: $^3$He depolarization rate measurements~\cite{Petukhov2010}, line 2: $^3$He magnetometer~\cite{chu2013}, line 3: $^3$He depolarization rate measurement~\cite{guigue2015}, line 4: $^{129}$Xe-$^{131}$Xe-Rb comagnetometer~\cite{bulatowicz2013,bulatowicz2013misc}, line 5: this work, line 6: $^3$He-$^{129}$Xe comagnetometer~\cite{tullney2013}. }
\end{figure}

The measured $X$ is connected with the monopole-dipole interaction parameters by~\footnote{See Sec.~II of the supplementary materials.}:
\begin{equation}~\label{eq:Xi}
\hbar X_i=v(r)\hat{\boldsymbol{r}}\cdot\hat{\boldsymbol{z}}\frac{\langle\sigma_{n}\rangle_i}{\langle I\rangle_i},
\end{equation}
where $I_i$ is the nuclear spin of $^i$Xe. Based on the nuclear spin analysis in Refs.~\cite{ressell1997,bulatowicz2013}, we have $\langle \sigma_n\rangle_{129}/\langle I\rangle_{129}=1.436$, and $\langle \sigma_n\rangle_{131}/\langle I\rangle_{131}=-0.303$ in Eq.~\eqref{eq:Xi}. These values lead to $\hbar(X_{129}-RX_{131})=0.41\langle v(r)\hat{\boldsymbol{r}}\cdot\hat{\boldsymbol{z}}\rangle$~\cite{miscR}, where the average symbol means the integration of $v(r)\hat{\boldsymbol{r}}\cdot\hat{\boldsymbol{z}}$ over the volumes of both the BGO mass and the atomic cell. Together with Eq.~\eqref{eq:Vr}, we can extract the upper limit of the coupling constant product $|g_s^Ng_p^n|$. Figure~\ref{fig:gsgn} shows the constraints on $|g_s^Ng_p^n|$ set by this work and other related experiments. An improved upper limit of the monopole-dipole coupling constants is achieved in the interaction range 0.11 - 0.55 mm, which corresponds to the propagator mass range of 0.36 - 1.80 meV. A maximum improvement factor of 30 is reached for the interaction length at 0.24~mm.

Current results are mainly limited by the Rb magnetometer sensitivity, and by the minimum distance between the BGO mass and Xe atoms. In order to further improve the search sensitivity, the multipass cavity technique can be implemented to increase the Rb magnetometer sensitivity~\cite{hao2021}. The wall separating Xe spins and the BGO mass can also be made thinner using microfabrication techniques~\cite{kitching2018}, thus reducing the minimum distance between the two. These efforts could improve the search sensitivity by several orders of magnitude in the sub-millimeter range.

A part of this work is carried out at the USTC Center for Micro and Nanoscale Research and Fabrication. We thank Prof. T. G. Walker, Prof. W. M. Snow, Dr. E. Smith, and Dr. H. Yan for helpful discussions. This work is supported by National Natural Science Foundation of China (Grant No.11774329) and the Strategic Priority Research Program, CAS (No. XDB21010200).


\begin{thebibliography}{40}
\bibitem [{bfm()}]{bfmisc}%
  \BibitemOpen
  \href@noop {} {}\bibinfo {note} {The gyro-compass effect due
  to the earth rotation is common for both BGO positions, therefore, it is
  cancelled in the differential measurement. Just for convenience, we choose
  the bias field to be in the east-west direction so that the gyro-compass
  effect on $R$ is minimized.}\BibitemShut {Stop}%
\end{thebibliography}
\end{document}